\documentclass[prd,twocolumn,superscriptaddress,showpacs,preprintnumbers,amsmath,amssymb,nofootinbib]{revtex4}
\usepackage{dcolumn}
\usepackage{bm}
\usepackage[latin1]{inputenc}
\usepackage[spanish,english]{babel}
\usepackage{amsfonts}
\usepackage{amssymb}
\usepackage{graphicx}

\newcommand{\be}{\begin{equation}}
\newcommand{\ee}{\end{equation}}
\newcommand{\bea}{\begin{eqnarray}}
\newcommand{\eea}{\end{eqnarray}}

\begin{document}
\title{{\bf Reply to ``Comment on `Insensitivity of Hawking radiation to an invariant Planck-scale cutoff' '' }}
\author{Ivan Agullo}
 \affiliation{ {\footnotesize Physics Department, University of
Wisconsin-Milwaukee, P.O.Box 413, Milwaukee, WI 53201 USA}}
\author{José Navarro-Salas}
\affiliation{ {\footnotesize Departamento de Física Teórica and
IFIC, Centro Mixto Universidad de Valencia-CSIC.
    Facultad de Física, Universidad de Valencia,
        Burjassot-46100, Valencia, Spain. }}

\author{Gonzalo J. Olmo}
\affiliation{{\footnotesize Instituto de Estructura de la Materia,
CSIC, Serrano 121, 28006 Madrid, Spain}}\affiliation{ {\footnotesize Physics Department, University of
Wisconsin-Milwaukee, P.O.Box 413, Milwaukee, WI 53201 USA}}

\author{Leonard Parker}
\affiliation{ {\footnotesize Physics Department, University of
Wisconsin-Milwaukee, P.O.Box 413, Milwaukee, WI 53201 USA}}

\date{April 14, 2010}

\begin{abstract}

We clarify the relationship between the conclusions of the previous Comment of A. Helfer \cite{comment} and that of our Brief Report \cite{briefreport}.

\end{abstract}


\maketitle

This Reply is intended to clarify our view of the trans-Planckian problem and how it differs from the, perhaps more conventional, view expressed in the previous Comment \cite{comment}.
In Hawking's original derivation of the thermal spectrum of radiation emitted by a black hole,
 which rests on the formalism of Bogolubov transformations,
the issue of trans-Planckian frequencies arises because an outgoing mode that reaches future null infinity at infinitely late retarded times will suffer a divergent blueshift when propagated backward in time to past null infinity in the rest frame of the black hole. Similar trans-Planckian energies also enter into the derivation of the acceleration radiation in terms of Bogolubov coefficients. 
Any outgoing Rindler mode corresponds to modes with  exponentially large frequencies at late times with respect to a fixed inertial observer.  
The fundamental point to address, in our view, is whether the conventional definition of trans-Planckian physics, as explained in \cite{comment}, really must enter into the derivations of the thermal Hawking and acceleration radiation. Is the trans-Planckian problem tied to these effects in an essential way, or is it an artifact of the mathematical formalism, as already suspected by many authors (see, for instance, \cite{strominger-polchinski})?\\
\indent In our brief report \cite{briefreport}, we argued that the analysis of the trans-Planckian problem for the acceleration radiation offers a new way to look at the trans-Planckian problem for Hawking radiation by a black hole.  The key point is that the analysis of the acceleration radiation using an Unruh-DeWitt particle detector involves only the invariantly defined proper time along the accelerated world line, so it is natural there to define the trans-Planckian region in terms of this proper time. When this idea is translated over to the black hole spacetime in \cite{briefreport}, it gives an invariant definition of the trans-Planckian region, corresponding to the narrow darkened region of Helfer's Fig. 2 \cite{comment}.  The underlying reason is that the response of the detector is characterized by the two-point correlation function.  Although the picture of propagation backward in time of the modes in the Hawking derivation would suggest that the gray region in that figure should characterize the trans-Planckian physics, the derivation in terms of the detector response function depends on a more narrow invariant definition of trans-Planckian physics. Our analysis in terms of two-point functions suggests that the Hawking effect is indeed a low-energy phenomenon.\\
\indent  Let us briefly rephrase our argument. The transition probability rate between two energy levels $E_i, E_f$ (an upper excited level $E_2$ and a lower one $E_1$) of an atomic detector interacting with a scalar field and undergoing uniformly accelerated motion (with acceleration $a$) is proportional to the response rate function
\be \label{RRF0}
{\dot F_{i \to f}(\Delta E)} = \int_{-\infty}^{+\infty}d\Delta
\tau e^{i (E_i -E_f) \Delta \tau}G_M(\Delta \tau -i\epsilon)  \ , \ee
where $G_M(\Delta \tau)= -\hbar (a/2)^2/(4\pi^2 \sinh^2[a/2(\Delta \tau -i\epsilon)])$ is the two-point function of the scalar field in the Minkowskian vacuum evaluated along the accelerated trajectory, and $\tau$ is the proper time along the trajectory ($\Delta \tau\equiv \tau_1-\tau_2$). The thermal response of the detector is obtained via the detailed balance relation $e^{-(E_2 - E_1)/T}= \dot F_{1\to 2} /\dot F_{2\to 1}$, from which one finds $T=a\hbar/2\pi$. In that approach trans-Planckian physics could appear in the ultra-short lapses of proper time involved in evaluating (\ref{RRF0}). Then, in order to probe the contribution of trans-Planckian physics to the thermal result, the natural thing is to examine the effect of a cut-off (of order of Planck scale) in the proper time lapse $\Delta \tau$. This corresponds to the  invariant cut-off introduced in \cite{briefreport}.\\
\indent On the other hand, one could perform the following change of variables in (\ref{RRF0}): $U\equiv t-x=-a^{-1} e^{- a \tau}$ ($t$ and $x$ are inertial coordinates and we are assuming that the acceleration of the detector is in the $x$ direction). The inertial two-point function now reads $G_M(U_1, U_2)= -\hbar/(4\pi^2 (U_1-U_2-i\epsilon)^2)$ and expression (\ref{RRF0}) corresponds then to expression (20) of our paper \cite{briefreport}.  Another possibility is to assert that trans-Planckian physics in that integral appears when differences in $U$ coordinates smaller than the Planck length $\ell_p$ are considered, that  is, when $(U_1-U_2)^2<\ell_{p}^2$. This region corresponds to the gray region of Fig. 2 of the Comment {\cite{comment}. However, it is clear in this context that the coordinates $U$ do not have any absolute meaning because there is no preferred inertial frame.  Therefore, in this case it seems more physical to characterize the trans-Planckian physics in terms of the invariant proper time lapse, by saying that trans-Planckian physics appears when  $\Delta \tau^2 < \ell_p^2$. This latter region can be re-expressed in terms of $U$ coordinates as $(U_1-U_2)^2< \ell^2_p a^2 U_1U_2/4\pi^2$ and corresponds to the darkened region of Fig. 2 of the Comment \cite{comment}. This expression has an invariant physical meaning as emphasized in our paper \cite{briefreport} (for instance, one can immediately check  that it is invariant under Lorentz boosts $U\to\gamma U$ of rapidity $\gamma$).\\
\indent  The next step is to separately evaluate the effects of eliminating each of the above regions ($(U_1-U_2)^2<\ell_{p}^2$ or $\Delta \tau^2 < \ell_p^2$) in the computation of the transition probabilities. 
However, expression (\ref{RRF0}) cannot be used to evaluate the effect of such a cut-off. The distributional character of the two-point function $G_M(\Delta \tau -i\epsilon)$, manifests itself in the usage of the $i\epsilon$ prescription, prevents the introduction of a cut-off in (\ref{RRF0}). The  $i\epsilon$ prescription is incompatible with the presence of a cut-off in the integral path  \cite{kay-wald}.  As sketched in our paper, one can bypass this situation by subtracting from the two-point function in the Minkowski vacuum the corresponding two-point function of the accelerated observer in the Rindler vacuum
\be \label{induced0}{\dot F}_{i \to f}(ind)=  \int_{-\infty}^{+\infty}d\Delta
\tau e^{i( E_i-E_f) \Delta \tau}[G_M(\Delta \tau)- G_A(\Delta \tau)] \ , \ee
where $G_A(x_1,x_2)\equiv \langle 0_A | \phi(x_1) \phi(x_2)|0_A \rangle$ and $|0_A\rangle$ is the usual Rindler vacuum. This subtraction makes the integrand a smooth function and the $i\epsilon$ can be eliminated. Therefore, one can properly estimate the contribution of trans-Planckian physics in the previous integral by introducing an appropriate cut-off.  Additionally, the subtraction of $G_A(\Delta \tau)$ has physical meaning because the resulting integral corresponds to the probability of induced  (or stimulated) absorption or emission of a quantum by the detector.
One then finds that, as pointed out in our original brief report, the contribution of the interval $\Delta \tau^2 < \ell_p^2$ to the above integral is negligible. Hence, we conclude that trans-Planckian lapses of proper time are not fundamental for obtaining the thermal result. On the contrary, if we repeat the computations using the (non-invariant) $U$ coordinates and we eliminate the  interval $(U_1-U_2)^2<\ell_{p}^2$ the thermal result gets totally modified.\\
\indent In summary, our argument shows that one can derive the acceleration radiation effect in a plausible way without invoking trans-Planckian physics. Our definition of trans-Planckian physics differs from the more standard definition used in Helfer's Comment and, as emphasized by Helfer, it can allow trans-Planckian ``precursors'' of the Rindler quanta from the point of view of a fixed inertial observer. However, these ``precursors'' are not detectable by an  inertial observer and their physical relevance is not clear. In fact, the inertial observer describes the excitation of the accelerated detector in an entirely different way than the accelerated observer. While the accelerated observer describes the excitation in terms of the absorption of Rindler quanta, the inertial observer describes the excitation as the emission of  Minkowski quanta \cite{unruh-wald}.  \\
\indent The same considerations can be applied to the Hawking radiation. In fact, as shown in our paper \cite{briefreport}, the mathematical expression giving the mean number of particles emitted per unit time by the black hole at late times is closely related to (\ref{induced0}), with the proper time $\tau$ replaced by the advanced time $u$ in the Schwarzschild geometry. One can better understand why this relation is so close by taking into account the fact that the induced transition probability of the detector is proportional to the energy density of the radiation, where the proportionality is given by one of the Einstein coefficients.
That implies that expression (\ref{induced0}) is precisely the mean number of particles present in the thermal bath of radiation detected by the accelerated observer times a factor $\Delta E/2\pi$. This is exactly the same expression that appears in the derivation of the Hawking effect (except for the factor $\Delta E/2\pi$) when computed using two-point functions (see \cite{briefreport} and references therein).  This strongly suggests that the invariant cut-off imposed for the accelerated detector corresponds in the black hole case to eliminating the region $\Delta u^2<\ell_p^2$ in the integral analogous to (\ref{induced0}). This corresponds again to the darkened region of Fig. 2 of the Comment \cite{comment}. The result one finds is that the black hole thermal spectrum is not sensitive to this type of trans-Planckian cut-off.\\
\indent We have shown how the clear physical picture offered by the  acceleration radiation effect strongly suggests that our new definition of trans-Planckian physics characterizes the physically significant region for the Hawking radiation as well. We believe that this characterization makes physically sense, as mentioned above, in spite of the fact that an analysis of the precursors of the Hawking quanta would involve trans-Planckian frequencies, as we already realized in our paper \cite{briefreport}.  
As in the acceleration radiation case, the problematic precursor modes may have no physical significance because they
are not detectable by an inertial observer in the distant past or a freely falling observer crossing the horizon of the black hole. Finally, we mention that the point of view offered in this note and in our brief report \cite{briefreport} is supported by the results from string theory where, in spite of the fact that one is using a quantum gravity theory, the prediction for the spectrum of black hole radiation is, surprisingly, unmodified at low energies. \\
\noindent { \bf Acknowledgements.}
We thank Professor Helfer for useful comments and enlightening discussions.

\end{document}